\documentclass[
 reprint, 
 amsmath,amssymb,
 aps,
]{revtex4-2}

\usepackage{graphicx}
\usepackage{dcolumn}
\usepackage{url}
\usepackage{bm}

\newcommand{\derivfrac}[2]{\frac{\text{d} #1}{\text{d} #2}}

\begin{document}

\preprint{APS/123-QED}

\title{Modeling Insights from COVID-19 Incidence Data: Part II - Why are compartment models so accurate?}


\author{Ryan Wilkinson}
\author{Marcus Roper}

\date{\today}

\begin{abstract}
The SIR-compartment model is among the simplest models that describe the spread of a disease through a population. The model makes the unrealistic assumption that the population through which the disease is spreading is well-mixed. Although real populations have heterogeneities in contacts not represented in the SIR model, it nevertheless well fits real U.S. state data at multiple points throughout the pandemic. Here we demonstrate mathematically how closely the simple continuous SIR model approximates a model which includes heterogeneous contacts, and provide insight onto how one can interpret parameters gleaned from regression in the context of heterogeneous dynamics.
\end{abstract}

\maketitle

\section{Introduction}\label{sec:Introduction}
Differences in rates of contact and of susceptibility between individuals strongly affect both their likelihood of catching COVID-19 and their health outcomes once infected. Incorporating these heterogeneities into models of disease spread is essential to understand the differential impacts of the disease upon different subpopulations, such as Black or Brown Americans \cite{reyes2020disproportional,macias2020COVID}, nursing home residents \cite{yourish2020one}, incarcerated \cite{saloner2020COVID} and unvaccinated individuals \cite{moghadas2021impact}. As well as ensuring a more equitable understanding of the disease, baseline questions about whether, e.g. it was safe to reopen schools in the middle of the pandemic, can not be definitively answered without considering the different levels of vulnerability of the communities affected: students, their caretakers, teachers and school staff. \\
\indent However, heterogeneity-capturing models contain many unknown parameters that are difficult to fit to real data, and are hard to interpret once fit. Accordingly, public health departments continue to make predictions about the progress of the epidemic and about the effectiveness of social distancing based on so-called well-mixed models (reviewed in \cite{bertozzi2020challenges}), and these models can be made to fit existing case data very well. They suffer, however, from having assumptions that are too simplistic to honestly reflect complex social behaviors inherent in disease spread, and as such, parameter fits should be interpreted with caution. In the well-known SIR model \cite{diekmann2000mathematical}, which is the focus of this paper, a group of susceptible individuals (\textit{S}) transition via contact with infectious individuals to the infectious (\textit{I}) group, and after some time recover or are removed (\textit{R}). The compartments evolve according to:
\begin{equation}\label{eq:Simple_SIR}
    \derivfrac{S}{t} = - \beta \frac{S I}{N_{tot}}~, \quad
    \derivfrac{I}{t} = \beta \frac{S I}{N_{tot}}  - \gamma I~,
\end{equation}
while $\frac{dR}{dt}=\gamma I$ ensures that, neglecting disease and other mortality, the total number of individuals $S + I + R = N_{tot}$ remains constant. The \textit{susceptibility} coefficient $\beta$ represents the number of infections caused by a single infected individual in an otherwise susceptible population in unit time. $\gamma$ is a basic recovery rate.
For the above equations to be valid the population should be well-mixed: everyone in the population interacts with everyone else at all times. Given geographic considerations and changes in mixing behavior during a pandemic, this assumption cannot be true. Accordingly, much modern epidemiological modeling, including of COVID-19 \cite{bertozzi2020challenges, arenas2020mathematical,tolles2020modeling}, has focused on the role of heterogeneous contacts, either by ramifying compartments or by using networks to model connections between individuals \cite{moreno2002epidemic, britton2020mathematical, li2020simulating, keeling2005networks,eames2002modeling,keeling2005implications,moreno2002epidemic,bansal2010dynamic,volz2007susceptible,yan2008distribution}. Although inclusion of heterogeneities can drastically affect e.g. thresholds for herd immunity \cite{britton2020mathematical}, the multiplication of parameters that occurs when heterogeneities are added to models makes it hard to validate these predictions against real data. Previous work has shown that under certain conditions heterogeneous models can be approximated by well-mixed models \cite{keeling1997correlation,sahneh2013generalized,bansal2007individual}, but there is limited data showing these conditions are met by real epidemics. \\
\indent In Part I of this paper, we showed that the COVID-case curves from different US states and during different surges can be clustered into between 4 and 6 groups. The collapse of case curves from different states to a small number of master curves suggests, that in spite of heterogeneous COVID transmission rates and impacts, relatively low model complexity is needed to reproduce the overall growth and the decay in number of infectious cases in a surge. Alignment of case data from different states involved translating data in time, and rescaling number of cases by a population size that was detected during data clustering (Part I), and that we interpreted to be the size of the subpopulation through which COVID-transmission was occurring. It follows that any mathematical model to describe this data collapse, must have the same symmetries; and these symmetries are present within the SIR model \cite{bertozzi2020challenges}.

\begin{figure}[!ht]
\includegraphics[width = \columnwidth]{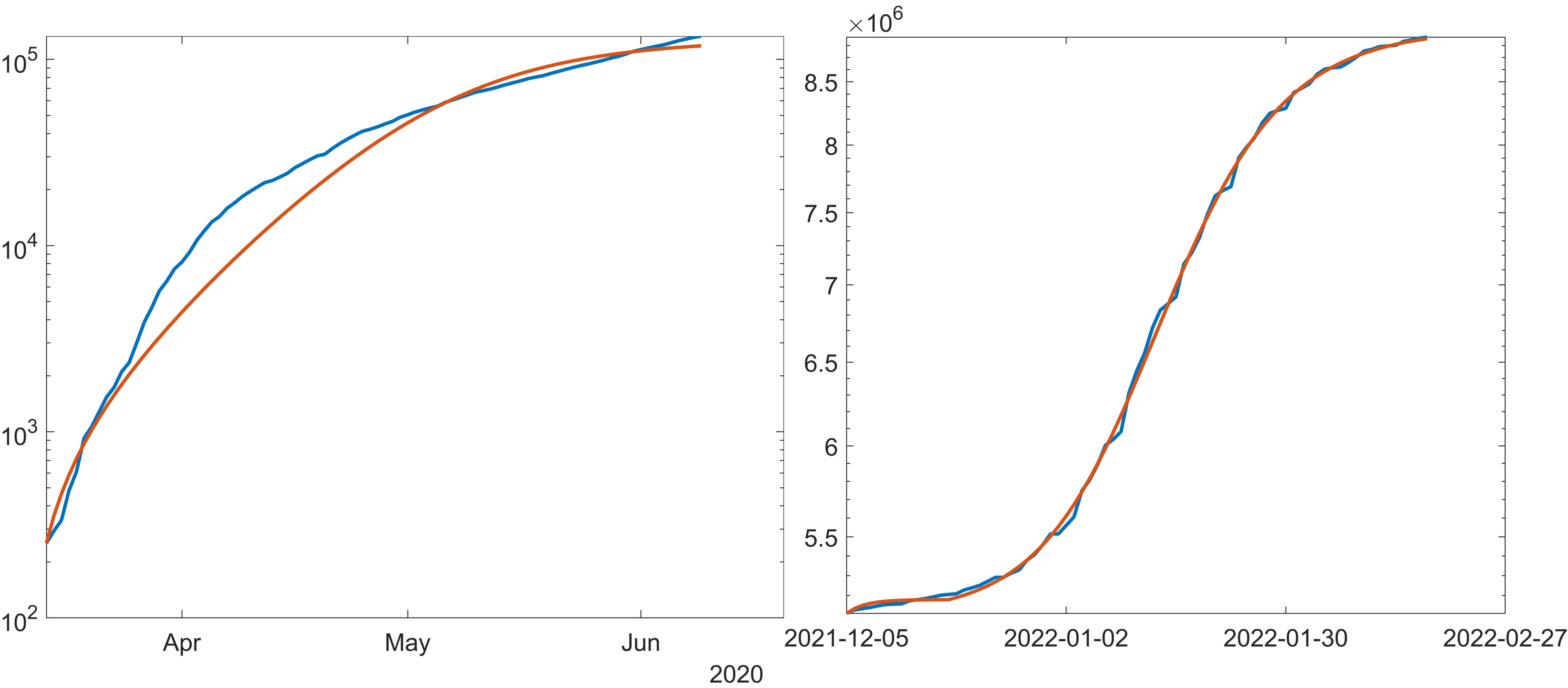}
\caption{\label{fig:CA_Showcase} California's cumulative COVID case data (blue) is reasonably approximated by the SIR model (orange) both in its initial stage and during the Omicron surge (left and right panels, respectively). Initial COVID data is highly dependent on testing capacity which was highly variable during the initial wave of COVID, so goodness of fit of the initial wave is best gauged by the fit to the latter part of the data.}
\end{figure}

\begin{figure*}[!ht]
    \includegraphics[scale = 0.5]{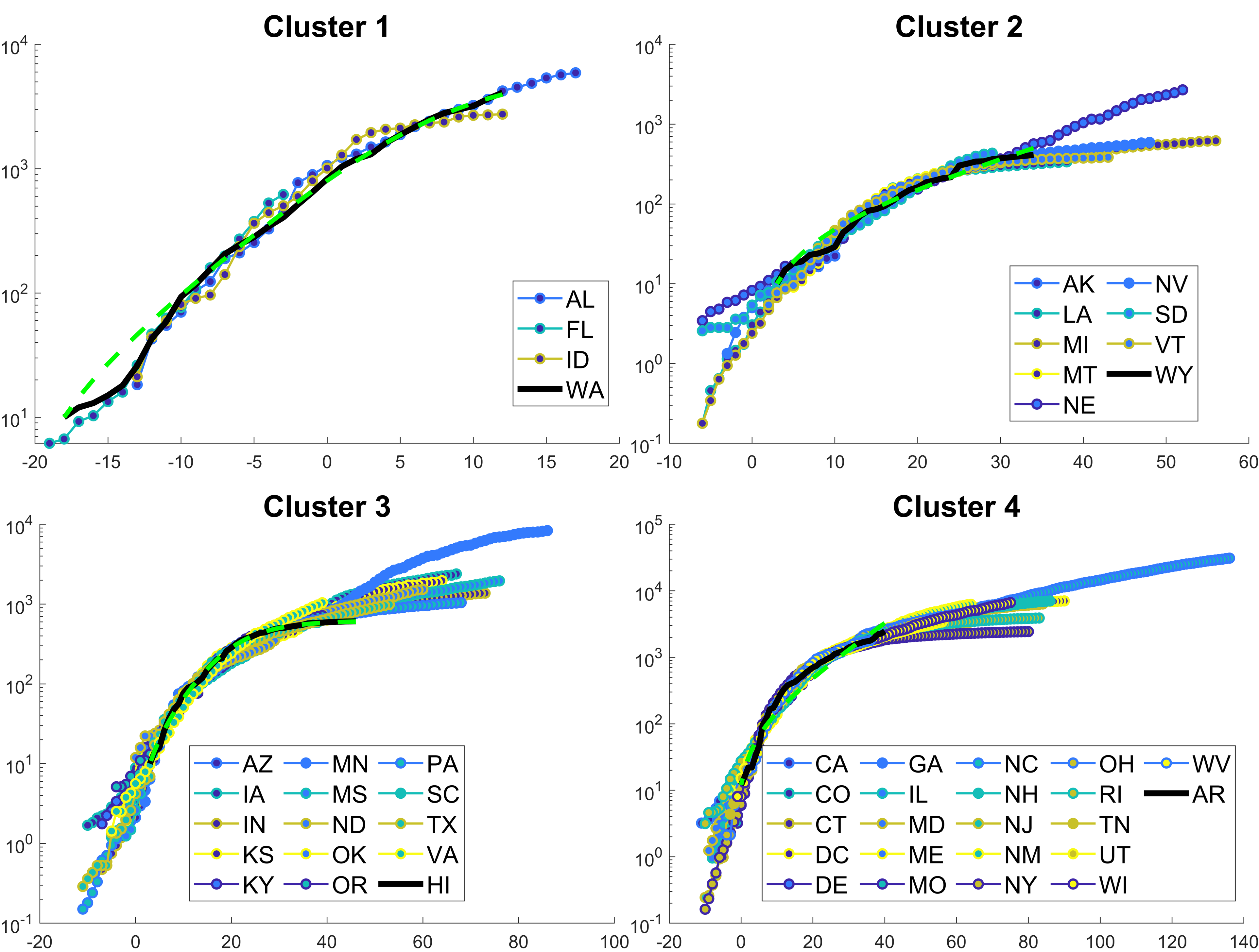}
    \caption{\label{fig:early_clusters}Shown are the computed clusters grouped together by the methods described in Part I. The least-fitting cluster (in the authors' opinion) lining up with Arkansas still fits reasonably well when the fit at later stages is considered, especially when one keeps in mind that case data is highly unreliable and sensitive to stochastic effects at the beginning of the pandemic.}
\end{figure*}

\begin{figure}[!ht]
\includegraphics[width = \columnwidth]{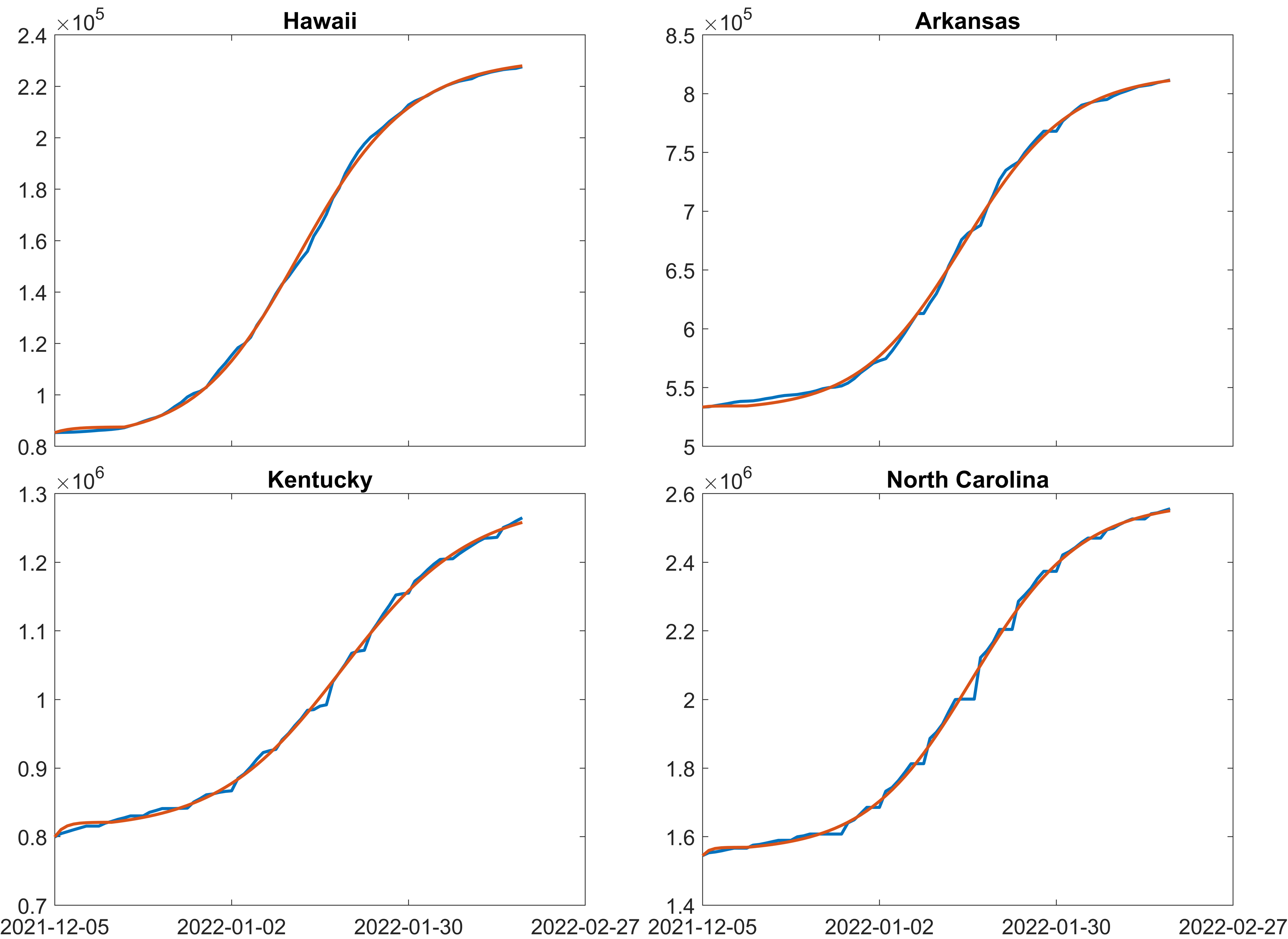}
\caption{\label{fig:OMC_Showcase} The SIR model (orange) fits Omicron surge data (blue) robustly for the vast majority of U.S. states. Shown are four examples of good fits for states with wildly different social/demographic structures and differing smoothness of data reporting.}
\end{figure}

We therefore postulated that the SIR model, described above, might be able to describe some of the families of case dynamics. We fit the SIR model to data from the first COVID surge, specifically the cumulative number of infections detected in each US states from the beginning of their respective outbreaks in February or March of 2020, up to May 20, 2020 (by which time stay at home orders had been relaxed, changing the transmission rate of the disease \cite{Repoening}).  In total, the SIR model fits the US state data for 2 of the 4 initial COVID phase clusters, accounting for 26 of the 52 states and territories in the data set (see Fig \ref{fig:CA_Showcase} and Fig \ref{fig:early_clusters}) of this paper for examples of early COVID SIR fits). The SIR model performed much better for the most recent surge in the Omicron variant with few exceptions, closely conforming to the true case trajectory. Representative fits from SIR models to the Omicron variant are shown in Fig \ref{fig:OMC_Showcase}.
In this paper, we examine two questions: First, given its neglect of heterogeneities in population contacts, why does the well-mixed SIR differential equation fit real pandemic data so well? Second, when we fit the SIR model to real observational data, can we interpret its fitted parameters, when the interaction processes that are parameterized are themselves not realistic? We find that the fitted $\beta$ parameter implicitly accounts for heterogeneous interaction dynamics, and the $N_{tot}$ parameter reveals an effective epidemic population size. We then apply our findings to the surprisingly well-fit omicron variant surge.\\

\section{Results}
\subsection{A Heterogeneous Extension}\label{sec:A Heterogeneous Extension}
To examine why the simple SIR model fits real pandemic data at all, we consider a multi-population SIR model that allows for populations to be heterogeneous in their contacts and susceptibilities. We then interrogate simulations resulting from this model to examine how important such heterogeneity is in the dynamics, and what fitting such a model with a homogeneous model means mathematically. \\
\indent The model assumes that the population is split into $K$ subpopulations, with population sizes $N_i$, $i = 1,2,\hdots,K$, that interact with each other with contact rate $\beta_{ij}$, i.e. $\beta_{i j}$ is the number of infections a single infected individual in subpopulation $i$ could cause in subpopulation $j$, if $j$ contains only susceptible individuals. 
Our subpopulations represent geographic or demographic partitions of a population of size $N_{tot}$. In this first treatment we neglect differences in recovery rate based on individual characteristics. The susceptible, infectious, and removed quantities of subpopulation $i$ is given by $S_i$, $I_i$, and $R_i$ respectively and also write $\sum_i S_i = S$,  $\sum_i I_i = I$, $\sum_i R_i = R$, and $\sum_i N_i = N_{tot}$. Consider the model:
\begin{equation}\label{eq:SIR_Alg_Manip}
\begin{aligned}
    \derivfrac{S}{t} 
    &= -\frac{\tilde{\beta}}{N_{tot}} S I - \sum_{i = 1}^K \sum_{j = 1}^K \big(\frac{\beta_{i j}}{N_j} - \frac{\tilde{\beta}}{N_{tot}}\big)S_i I_j  \\
    \derivfrac{I}{t}
    &= \frac{\tilde{\beta}}{N_{tot}} S I  - \gamma I + \sum_{i = 1}^K \sum_{j = 1}^K \big(\frac{\beta_{i j}}{N_j} - \frac{\tilde{\beta}}{N_{tot}}\big)S_i I_j
\end{aligned}
\end{equation}
To emphasize the similarity between the multi-population and the single-population SIR model we write the equation in a form with a simple SIR part, with arbitrary coefficient $\tilde{\beta}$ plus a residual. We obtain a type of best SIR fit by minimizing the $L^2$ norm of the ratio of the heterogeneous residual to the homogeneous term on an arbitrary interval $(t_1,t_2)$, i.e. we minimize the size of the function 
\begin{equation}\label{eq:F}
    F(\tilde{\beta},t)
\equiv
\frac{
\sum_{i = 1}^K \sum_{j = 1}^K \big(\frac{\beta_{i j}}{N_j} - \frac{\tilde{\beta}}{N_{tot}}\big)S_i I_j
}{
\frac{\tilde{\beta}}{N_{tot}}S I
}   
\end{equation}
This yields the unique minimum
\begin{equation}\label{eq:beta_tilde}
\begin{aligned}
    \tilde{\beta} &= N_{tot}\frac{
    \int_{t_1}^{t_2} \frac{
    \Big(\sum_{i = 1}^K \sum_{j = 1}^K \frac{\beta_{i j}}{N_j} S_i I_j\Big)^2
    }{
    \big(SI\big)^2
    }\ dt
    }{
    \int_{t_1}^{t_2} \frac{
    \sum_{i = 1}^K \sum_{j = 1}^K\frac{\beta_{i j}}{N_j} S_i I_j 
    }{
    SI
    }
    \ dt 
    }
\end{aligned}
\end{equation}
We explore this quantity for a certain class of subpopulation interaction structure in Section (\ref{sec:Intermediate Mixed/Unmixed Models}) and (\ref{sec:A Model with Network Structure}).
\subsection{Intermediate Mixed/Unmixed Models}\label{sec:Intermediate Mixed/Unmixed Models}
\begin{figure*}[!ht]
\includegraphics[width = \textwidth]{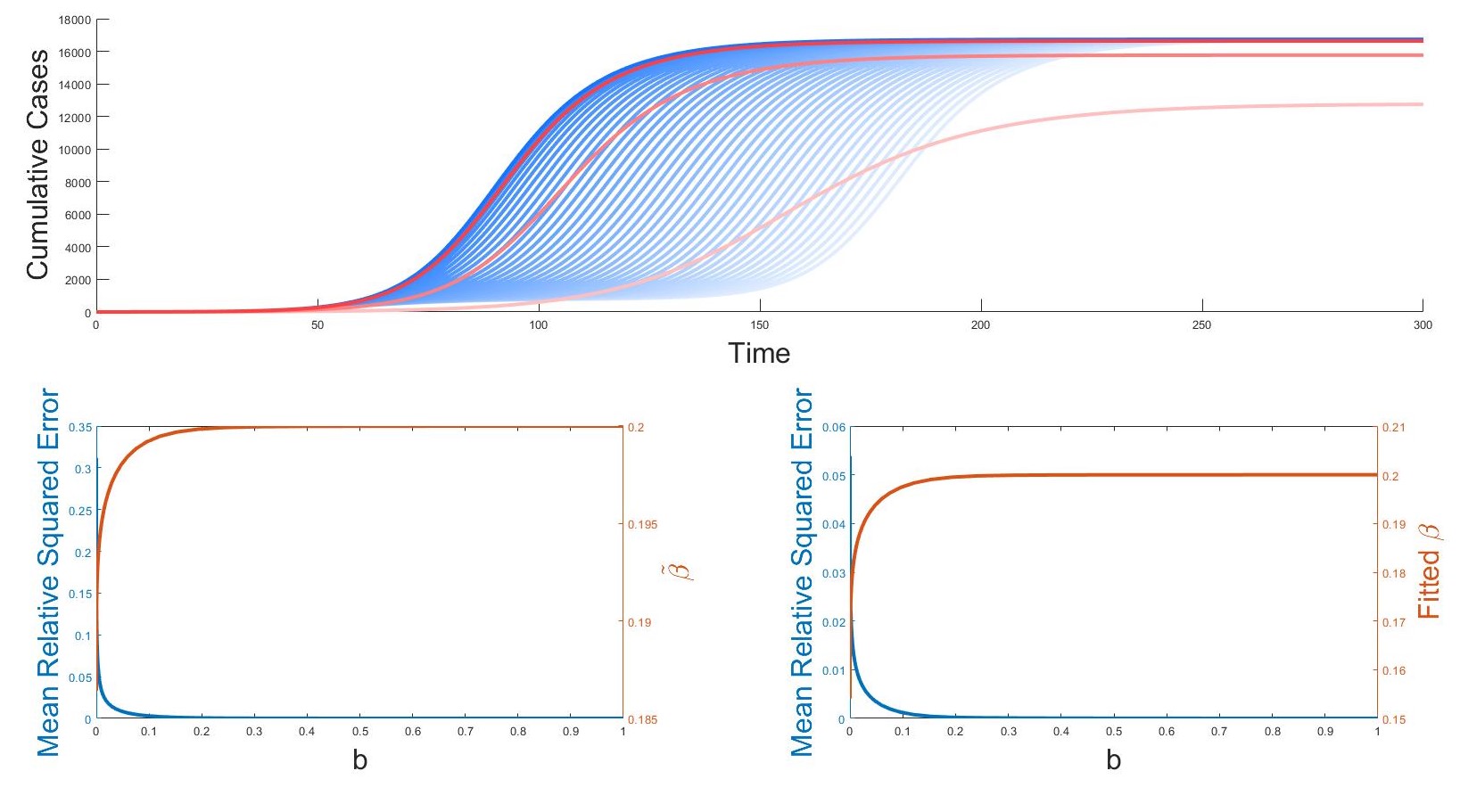}
\caption{\label{fig:Model_Transition} A single well-mixed population model can represent aggregate spread of disease through a linked set of subpopulations, even with relatively weak inter-subpopulation mixing. 
Top: Cumulative cases from the model in Eq. (\ref{eq:SIR_Alg_Manip}) with $\beta_{i j}$ given by Eq. (\ref{eq:Model_Consistent_Contact_Matrix_ER_Graph}). Blue curves: subpopulation model with $b$ increasing from $b = 10^{-5}$ (lightest) to $b = 1$ (darkest) with log-spaced values. Red curves: optimal SIR fits, with $b=10^{-5}$ (lightest), $b \approx 10^{-2}$, and $b \approx 10^{-1}$ (darkest) cases. Bottom panels $\tilde{\beta}$ given by Eq. (\ref{eq:beta_tilde}) (left, orange) and by linear regression (right, orange) as a function of $b$ (orange), and the relative mean squared error of the true model relative to a simple SIR model using $\tilde{\beta}$ and $N_{tot}$ as a function of $b$ (blue). The plots included in this figure are from numerical simulations done with the parameters $\beta = 0.2$, $\gamma = 0.1$, $N_i = 1000$, and $K = 5$, and initial data $I_1(0) = 1$, $I_i(0) = 0 \forall \ i > 1$ and $R_i(0) = 0\, \forall i$. Both the $\tilde{\beta}$ and the regression calculations are done over an entire pandemic period (250 days). }
\end{figure*}

We first consider the case where subpopulations either do not interact or else interact at identical rates:
\begin{equation}\label{eq:Model_Consistent_Contact_Matrix_ER_Graph}
    \beta_{i j} = \beta\Big(\big(b\frac{N_j}{N_{tot}} + (1 - b)\big) \delta_{i j} + b \frac{N_j}{N_{tot}}(1 - \delta_{i j})A_{i j}\Big)
\end{equation}
Here $\delta_{i j}$ is the Kronecker delta, $b$ a mixing parameter ranging from $0$ to $1$, and $A_{i j}$ is an adjacency matrix describing which subpopulations interact. Assume initially $A_{i j} \equiv 1$. When $b$ is $0$ the populations do not mix, and infections spread within but not between subpopulations, and when $b = 1$ the subpopulations mix completely, effectively merging into a single homogeneous population by construction of Eq. (\ref{eq:Model_Consistent_Contact_Matrix_ER_Graph}). Scaling $\beta_{i j}$ by $N_j$ ensures that interactions between subpopulations are proportionate to their sizes. \\
\indent A homogeneous SIR model fits this model for $b \gtrsim 0.2$ (Fig (\ref{fig:Model_Transition})). Moreover, the $\tilde{\beta}$ gathered by Eq. (\ref{eq:beta_tilde}) is very close to the $\beta$ achieved by least squares model fitting to the full simulation, although unsurprisingly least squares fitting performs better at fitting the actual simulated curve for lower values of $b$. The agreement between $\tilde{\beta}$ and the fitted $\beta$ is encouraging: for the consideration of real data, one can only deduce model parameters via some sort of fitting algorithm without knowledge of subcompartmental dynamics, but the result suggests that least squares fitting  optimally estimates the susceptibility parameter to maximize the simple SIR part of the model relative to the interpopulation dynamics.\\

\subsection{A Model with Network Structure}\label{sec:A Model with Network Structure}
\begin{figure}[!ht]
    \includegraphics[width = \columnwidth]{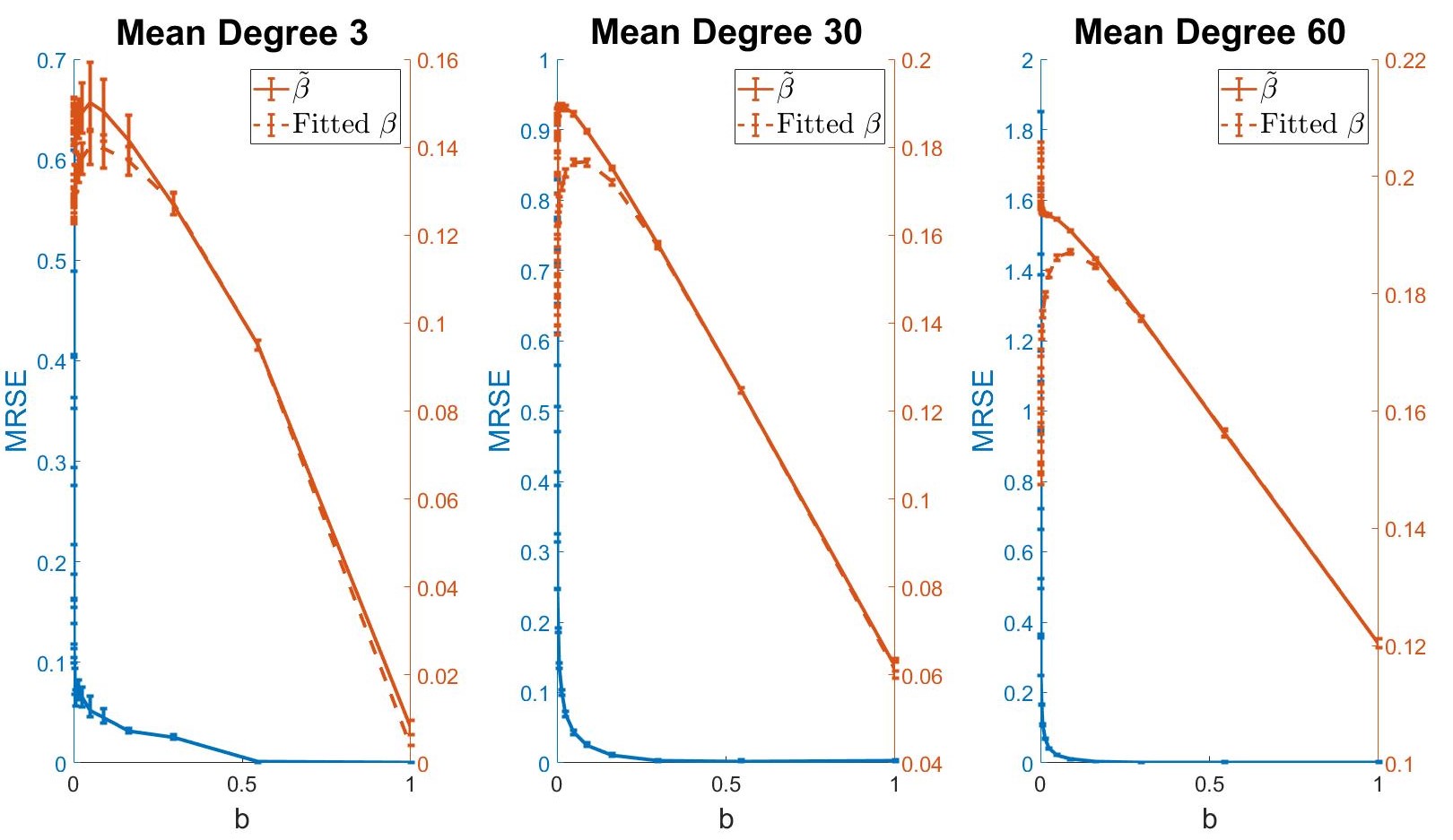}
    \caption{\label{fig:Graph_Dynamics} Networked subpopulations still allows for fitting by an SIR model. Erd{\H o}s-R{\'e}yni networks were simulated for with three different mean degree parameterizations. 100 replica simulations were run with $\beta = 0.2$, $\gamma = 0.1$, $N_i = 1000$, and $K = 101$. Shown: Close to identical estimates are obtained for $\tilde{\beta}$ using Eq. (\ref{eq:beta_tilde}) (orange, solid) or least squares fitting (orange, dashed). Mean Relative Squared error is shown for Eq. (\ref{eq:beta_tilde}) (blue, right axes).}
\end{figure}

To investigate possible effects of subcommunity interaction structure on the homogeneous-like dynamics, we now model subpopulations whose interactions are prescribed by a network with adjacency matrix $A_{i j}$. $A_{ij}=1$ indicates two subpopulations that maintain frequent contact with one another, such as a pair of communities which go to the same grocery store or school. For the purpose of analysis, we modeled random connections between subpopulations as random Erd{\H o}s-R{\'e}yni networks parameterized by mean degree. The SIR model approximates the graph dynamics model above $b = 0.2$, with the fit improving as the mean degree of each node is increased (Fig (\ref{fig:Graph_Dynamics})). Just as for a complete graph, $\tilde{\beta}$ agrees with $\beta$ from least squares fits. Narrowing of error bars as mean degree of the random network increases express the decreasing importance of network structure as the network becomes more densely connected.\\
\indent However, $\tilde{\beta}$ does not asymptote to $\beta = 0.2$ whether we use Eq. (\ref{eq:beta_tilde}) or least squares fitting. Instead, increasing mixing allows the model to gain awareness of how sparse the connections between subpopulations are, ultimately causing $\tilde{\beta}$ to decrease. Fig (\ref{fig:Graph_Dynamics}) suggests that as the mean degree increases, we expect that rate of decrease to slow, and that $\tilde{\beta}$ will eventually asymptote to $0.2$ as the network becomes complete (Fig (\ref{fig:Model_Transition})). Indeed, when all subpopulations have the same size, $\tilde{\beta}$ may be written as
\begin{widetext}
\begin{equation}\label{eq:beta_tilde_asymp}
    \tilde{\beta} = \beta \frac{
    \int_{t_1}^{t_2} \Big(\frac{K - (K-1)b}{K} \big(1 + K \sum_{i = 1}^K \frac{S_i}{S} (\frac{I_i}{I} - \frac{1}{K}) \big) + b \frac{\sum_{i = 1}^K\sum_{j=1}^K A_{i j} S_i I_j}{SI}\Big)^2\ dt
    }{
    \int_{t_1}^{t_2}\frac{K - (K-1)b}{K} \big(1 + K \sum_{i = 1}^K \frac{S_i}{S} (\frac{I_i}{I} - \frac{1}{K})\big)  + b \frac{\sum_{i = 1}^K\sum_{j=1}^K A_{i j} S_i I_j}{SI} \ dt  
    }    
\end{equation}
\end{widetext}
Both numerator and denominator include a term which relies on graph structure and gains weight with the mixing parameter $b$ and the covariance between the fraction of susceptibles and infecteds across subpopulations. The $\sum_{i = 1}^K \frac{S_i}{S} (\frac{I_i}{I} - \frac{1}{K})$ term represents a covariance measurement of the susceptible and infected percentage among the subpopulations, and is expected to be negative since, heuristically speaking, an increase in the number of infecteds corresponds to a decrease in the number of susceptibles, at least during a surge in cases. The $\frac{\sum_{i = 1}^K\sum_{j=1}^K A_{i j} S_i I_j}{SI}$ term encodes $\tilde{\beta}$'s dependence on subpopulation interaction structure. For a complete network, this term is 1, and it decreases to 0 as the network becomes more sparse.\\
\indent In the complete graph case and the case with network structure, we observe that $\tilde{\beta}$ is dampened by heterogeneous community contacts. Given $\tilde{\beta}$'s correspondence with the $\beta$ gleaned from data fitting, we now assume that the SIR parameter fit for $\beta$ actually underestimate the true person-to-person contact rate.

\subsection{Comparison with other estimators of disease spread}\label{sec:Connections to other Macroscopic Spread Estimators}

\begin{figure}[!ht]
    \centering
    \vspace{20pt}
    \includegraphics[width=\columnwidth]{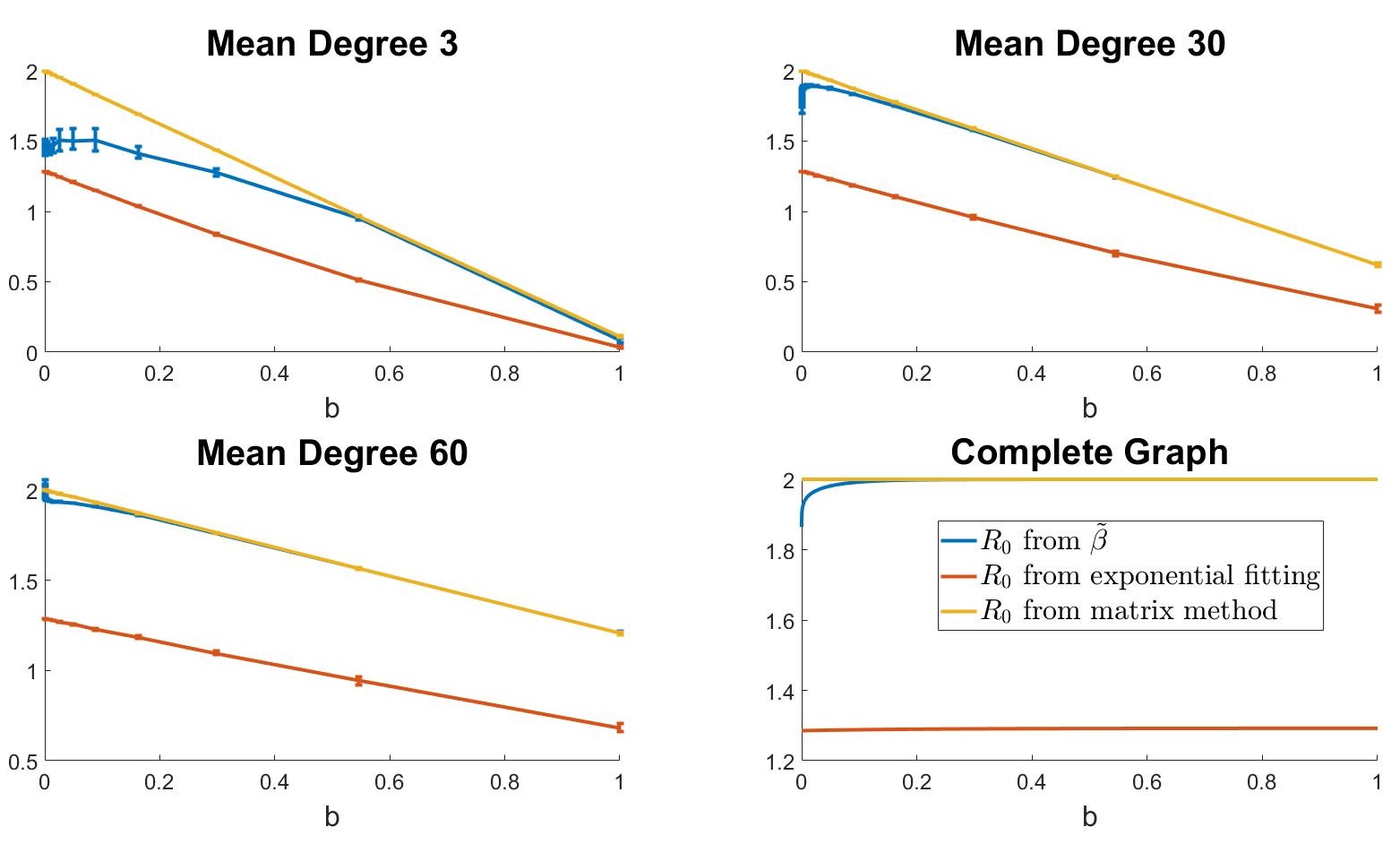}
    \caption{\label{fig:R0_Comparison} $R_0$ estimates from fitting entire case data curve optimally to well-mixed model (blue) agree with linearized analysis by next-generation matrix method (yellow) \cite{van2002reproduction}, but not to empirical fits to the data assuming exponential growth (red).}
\end{figure}

We have shown that one can generate a single population-level transmission rate, $\beta$, for a heterogeneous population. From $\beta$ we may derive the basic reproduction number $R_0=\beta/\gamma$, the expected number of secondary cases produced in a completely susceptible population, by a typical infectious individual\cite{diekmann1990definition}. Our estimation method relies on minimizing the error between heterogeneous and an SIR model over the entire time course of a surge through Eq. (\ref{eq:beta_tilde}). Prior estimation methods rely on the fact that in an SIR model, the parameter $\beta$ describes the linearized, or initial, exponential rate of growth of the number of infectious individuals. For our model, we estimate this linearized rate of growth by two methods: 1. fitting an exponential on the first 20 days of the simulation, and 2. the next-generation matrix method from \cite{van2002reproduction} (Fig (\ref{fig:R0_Comparison})).\\
\indent In \cite{van2002reproduction}, $R_0$ is computed for a general compartment model from the Jacobian matrix of the system. This matrix is evaluated at a disease-free equilibrium to determine the average number of individuals that a typical infectious person infects when the population is asymptotically disease free, i.e. $S_i = N_i$ for each $i$. Applying the method in \cite{van2002reproduction} to Eq. (\ref{eq:Model_Consistent_Contact_Matrix_ER_Graph}) and making the assumption that $N_i = N_j$ for all $i$ and $j$ as in the simulations, we obtain: 
\begin{equation}\label{eq:R0}
    R_0 = \frac{\beta}{\gamma}\Big( 1 - b + \frac{b}{K}\big(1 + \rho(A_{i j})  \big) \Big)
\end{equation}
where $\rho$ denotes the spectral radius of the matrix $A_{i j}$. In the complete network case (as in Fig (\ref{fig:Model_Transition})), $\rho(A_{i j}) = K - 1$, which yields $R_0 = \beta/\gamma$, matching the asymptote in Fig (\ref{fig:R0_Comparison}). \\

\indent As the mean degree of our random network increases, $R_0$ values computed from the next generation matrix method in \cite{van2002reproduction} and the $R_0$ computed from $\tilde{\beta}$ converge (Fig \ref{fig:R0_Comparison}). On all levels of network connectedness, and mixing parameter, exponential fitting consistently underestimates the contact rate. Under-estimation results from initial slowing in early growth, due to transmission being slower between subpopulations than within them. Conversely, for small values of the mixing parameter $b$, the next generation matrix method $R_0$ exceeds the optimal estimate from $\tilde{\beta}$, but the two estimators converge consistently at $b$ values between 0.1 and 0.45, depending on the mean degree of connectedness between subcompartments. \\

\indent The value of $R_0$ computed in \cite{van2002reproduction} comes from the linearized dynamics; for example it provides a threshold for the stability of disease-free equilibria (see \cite{hethcote2000mathematics}). By contrast the estimate for $R_0$ computed in this paper is computed by approximating the spread of the disease by a homogeneous model. Surprisingly, the two methods produce confluent results even under modest levels of mixing between subpopulations.\\

In contrast to constant-parameter growth fitting, time series fitting calculates the time-varying rate of exponential growth of the number of COVID cases, and thence infers the number of new infections caused by each COVID case. Unlike the SIR model, $R_0$, now called simply $R$, is not a constant, but typically varies over the course of a surge, and reflects not just the linearized dynamics of disease spread if an infected individual were transplanted to a population containing only susceptibles, but an actual estimate of new infections. Both SIR model and data fitting produce case curves that agree well with the surge of Omicron-variant cases, for which we use California as a representative example (section \ref{sec:Parameter_Conclusions}). However, the SIR model achieves this fit by assuming piecewise constant $\beta$, with a single ($\beta$, $N_{tot}$) pair covering most of the surge. We show in the next section that it is possible to forecast the end of the surge after an inflection point in a surge, and the total number of cases it will cause. Under the model, the surge ends only when the disease has been transmitted through an entire well-mixed subpopulation of size $N_{tot}$. By contrast, the time-varying $R$ value inferred by Epiforecasts \cite{Epiforecast_CA} gives a compelling visualization of how transmission rates decrease during the surge. However, data fitting can not distinguish between decreases in transmission rate due to inevitable decrease of number of susceptibles around each infectious individual, or due to public health orders changing the course of the epidemic. The success of the SIR model in fitting the data weights the first factor over the second, though does not discount the effectiveness of public health measures implemented at or before the start of the surge in controlling its trajectory.

\subsection{The $N_{tot}$ Parameter as Effective Case Surge Size} 

In the derivation of the SIR model,  $N_{tot}$ is the size of the population through which the disease is being transmitted. When using the model for data fitting, $N_{tot}$ is often treated as the size of the population; e.g. state or country from which the data was sourced \cite{bertozzi2020challenges}. However, real COVID cases occur in hot spots, and may not involve every individual in the studied population. In terms of our representation of this studied population by linked compartments, linkages between some compartments may be so weak that cases in one do not lead to a number of cases in the second that does not scale with the second compartment size. For this reason, we take advantage of the flexibility within our model of allowing $N_{tot}$ be fit alongside $\beta$.

Much like how we must reconsider the fitted $\beta$ in the context of heterogeneities, we must now reconsider what the fitted $N_{tot}$ represents. The SIR model believes that $N_{tot}$ gives the scale of the population modeled. In turn, our fitted $N_{tot}$ parameter tells us something new and useful: the epidemic population scale that the data is conveying. Such a fitted $N_{tot}$ can then be a new, data-driven scale by which we can estimate per-capita case numbers in the context of a given surge.

\begin{figure}[!ht]
\includegraphics[width = \columnwidth]{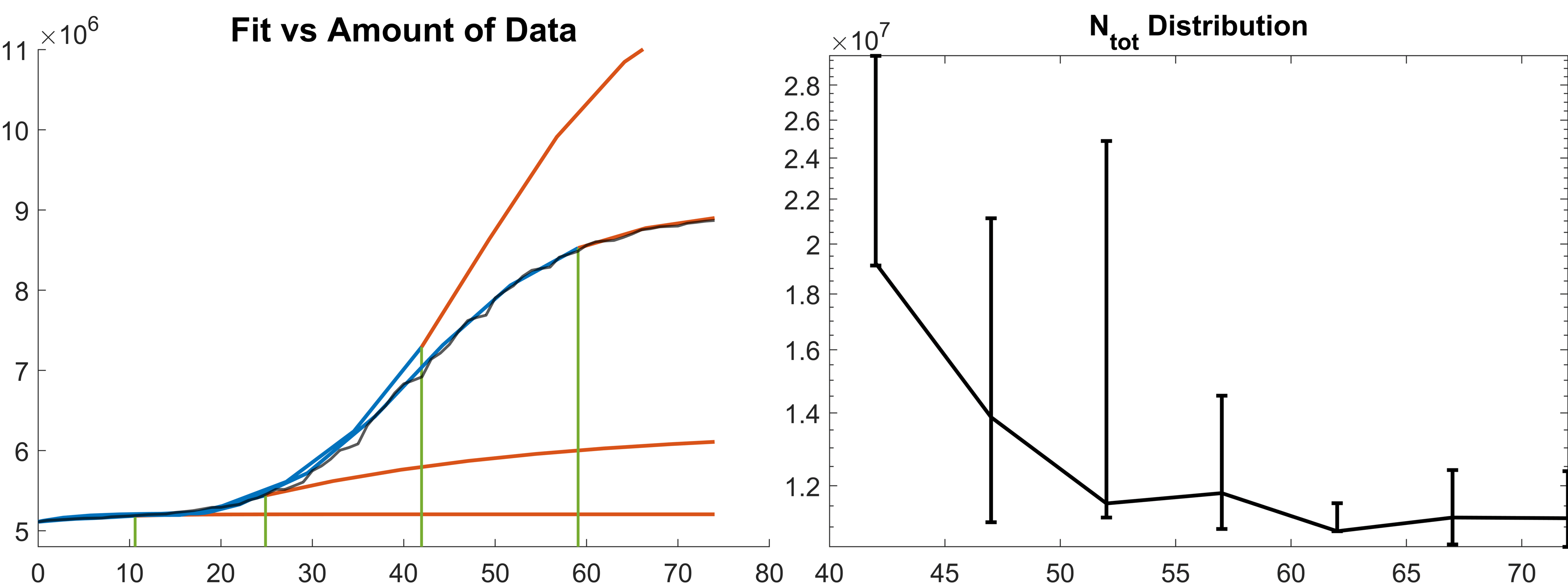}
\caption{\label{fig:Ntot_Fig} Fitting analysis of the $N_{tot}$ parameter for California's cumulative COVID case data during the Omicron variant surge. Left: the SIR model's best prediction (orange part of curve) when given a four different portions of the entire surge data (blue part of curve). Vertical green lines indicate the point where the SIR model starts predicting. As more data is given, the model better matches the actual cumulative case data (translucent black line). Right: the distribution of fitted $N_{tot}$ parameter values as a function of number of days of data given (horizontal axis in plot). Fit distributions are generated by taking the negative exponential of mean squared residual of the fit after randomizing the initial parameter guesses (see section SID). The distribution narrows around a more unique guess as more data is given.}
\end{figure}

Fitting $N_{tot}$, however, is surprisingly tricky. In the SIR model, $N_{tot}$ always appears coupled in a fraction with $\beta$, and as such, when the scale is not necessarily apparent from the data, only the ratio $\frac{\beta}{N_{tot}}$ can be tuned to fit. The scale of the initial conditions used to fit the data are also unreliable, in part due to a large degree of uncertainty of what the initial conditions are given cumulative case data, i.e. we have little information about how many people can be considered susceptible or removed at any given time in order to seed the fit with an accurate initial condition. Indeed, when not enough data to determine scale is present, the fits for $\beta$ and $N_{tot}$ are highly sensitive to the initial parameter estimates in our fitting algorithm (Fig \ref{fig:Ntot_Fig}).

We analyzed omicron surges (which are fit quite well by the SIR model, as explored in the next section) to determine exactly how much data is needed for scale to be fit robustly (Fig \ref{fig:Ntot_Fig}). We observe that predictive power and parameter fit robustness rapidly increase immediately after the inflection point in surge data, suggesting that the model needs to be sure of a future plateau of cumulative cases to truly determine scale independently of the ratio $\frac{\beta}{N_{tot}}$. This observation is supported mathematically by an analytical result on the eventual number of susceptibles as a function of model parameters:
\begin{equation}\label{eq:SIR_Infty}
    S_\infty e^{-\frac{\beta}{\gamma N_{tot}}S_\infty} = S(0) e^{-\frac{\beta}{\gamma N_{tot}}\big(N_{tot}-R(0)\big)}
\end{equation}
We see here a second indicator of scale for the SIR model present in the terminal dynamics of the model's trajectories. We reason that the inflection point in a case surge is the point after which these terminal dynamics become determined. In turn, the $N_{tot}$ parameter reveals itself uniquely.

\subsection{SIR Curve Fits and Parameter Conclusions from U.S. State Omicron Case Surge Data}\label{sec:Parameter_Conclusions}

Armed with the more nuanced interpretations of what fitted $\beta$ a $N_{tot}$ mean, we sought to draw conclusions from U.S. state data from the particularly infectious Omicron variant as it first rose to dominance among infectious individuals. Despite the relatively wide coverage of vaccines, the Omicron variant infected individuals as if they were purely susceptible, although vaccinated individuals were far less likely to end up hospitalized \cite{moghadas2021impact}. This high infection rate paired with the timing of two major U.S. holidays (Thanksgiving and Christmas) almost invariably led to homogeneous SIR-like dynamics for entire states (Fig \ref{fig:OMC_Showcase}), despite vast differences in population density and demographic structure. In fact, the only states not to exhibit clear homogeneous SIR dynamics appear to have extremely low-quality data. Given our early interpretations utilizing $\tilde{\beta}$ (Eq. \ref{eq:beta_tilde}), we conclude that some level of subpopulation interaction is happening state by state, with well-mixed dynamics occurring in the different subpopulations. Since the data considered is a surge of cases, we also expect the fitted $\beta$ to represent the bulk susceptibility parameter dampened by subpopulation interaction effects due to negative covariance between susceptibility numbers and infectious numbers (Eq. \ref{eq:beta_tilde_asymp}). Our parameter fits for $\beta$ in the following, therefore, are \textit{lower bounds} for the true person-to-person susceptibility rate rather than actual estimates.

We chose to use California as a case study for our parameter fits due to the absolute certainty of there being at least two loosely connected population centers for COVID spread (the San Francisco area and the Los Angeles area, separated by approximately 400 miles). Much like the vast majority of states, California's Omicron surge is extremely well approximated by a simple SIR curve. The most frequently-occurring set of parameters that fit the data put the basic reproduction number at approximately $R_0 \approx 3$ (a lower bound just as $\beta$ is), and $N_{tot} \approx 12 \ \text{million}$. The SIR model therefore thinks that California's data is actually represented by homogeneous spread among 12 million individuals. Our earlier interpretations temper this conclusion with the reality that the spread cannot be homogeneous, and leads us to the more nuanced conclusion: the spread of Omicron in California is dominated by the well-mixed disease spread amongst 1 or more distinct subpopulations totaling to approximately 12 million individuals, around 30\% of California's population. Among these communities, spread has a very rapid rate of at least $R_0 = 3$. This highlights the dominant role that a closely interacting set of individuals can play in driving pandemic case trajectories.

\section{Discussion}\label{sec:Discussion}

In the Part I of this paper we showed that COVID case growth curves from different US states can be collapsed onto a small number of universal curves, after translation in time and rescaling of the total number of cases. These symmetries are present in the SIR model, and indeed we find, alongside previous works \cite{bertozzi2020challenges,10.1371/journal.pone.0265815} that SIR models can well fit the initial COVID surges in 26 of 52 states and territories in our data set, although our SIR model fits use the simplest, constant-parameter ODE version. 

\indent Our analysis of data from states and territories provide empirical evidence that there is enough mixing between the disease hotspots within each state or territory that approximation by a well-mixed model is appropriate. The concordance between $\beta$ obtained by curve fitting and Eq. (\ref{eq:beta_tilde}) affirms that least squares fitting extracts information on the subpopulation interaction structure and dynamical asymmetry between subpopulations, and highlights the SIR model as a coarse-grained model for disease transmission in heterogeneous populations.  Remarkably, under moderate levels of transmission between subpopulations, our model-fitting approach, which is based on fitting the entire disease case curve, produces parameter estimates that agree well with the next generation matrix model based only on the linearized dynamics that include information only on initial epidemic features. We additionally gain the ability to reinterpret the total epidemic size $N_{tot}$ as a new, data-driven normalization factor. We may therefore construct case rates with respect to surge size, effectively considering normalized cases only among the population actively transmitting.\\

In general, simplicity of fitting to real data makes SIR (and similar well-mixed models, such as SEIR) powerful tools for predicting the ongoing course of an epidemic. However, the assumptions behind the model drastically simplify real patterns of human interaction and disease transmission, and the question of estimating community interaction structure remains. Indeed, an important corollary of our analysis is that, with even modest levels of connectedness between the subpopulations, they can function as well-mixed, making the details of the substructure undetectable by a well-mixed model. Although heterogeneity-capturing models aim to render these interactions, they introduce additional parameters that often must also be fit to the data.  It can be hard to distinguish improvements in fitting due to greater model realism from improvements due to increased parametric flexibility. 

Data fitted analyses make no assumptions about the underlying patterns of interaction, but create short term predictions based on fitting evolving exponential growth curves to the data. SIR models do not fit every phase of the case data; for example we could not fit the second surge identified in Part I, likely because of the complicating effects of time evolving contact rates caused by changing work patterns and public health orders, and the presence of previously recovered individuals within the population. By contrast, although models of the the Omicron surge encounter similar complications, reluctance to reimpose social distancing measures, and the ability of Omicron to readily infect vaccinated and previously-infected individuals lead to a time course that is closer to the first surge, and that can be well-fit by an SIR model.


\bibliography{B}

\end{document}